\documentclass [12pt,amsmath, amssymb, aps, nofootinbib, notitlepage] {article}
 \usepackage{lipsum}
\usepackage[margin=1.75cm]{geometry}
\usepackage{amssymb}
\usepackage{graphicx}
\usepackage{amsmath}
\usepackage[hidelinks]{hyperref}
\usepackage{array}
\usepackage{verbatim} 
\usepackage{multirow}
\usepackage{manfnt}
\usepackage{float}
\usepackage{caption}
\usepackage{braket}
\usepackage[numbers]{natbib}

\numberwithin{equation}{section}

\usepackage{tocloft}

\def\be{\begin{equation}}
\def\ee{\end{equation}}

\usepackage{authblk}
\usepackage{color}

\title {The complex timeless emergence of time in quantum gravity}
\author {Daniele Oriti}
	\affil{Arnold Sommerfeld Center for Theoretical Physics, \\ Ludwig-Maximilians-Universit\"at München \\ Theresienstrasse 37, 80333 M\"unchen, Germany \\ daniele.oriti@physik.lmu.de}

 \date{}

\begin{document}

\maketitle

\begin{abstract}
We present the arguments suggesting that time is emergent in quantum gravity and discuss extensively, but without any technical detail, the many aspects that can be involved in such emergence. We refer to both the physical issues that need to be tackled, by quantum gravity formalisms, to realize concretely this emergent picture of time, and the conceptual challenges that have to be addressed in parallel to achieve a proper understanding of it.
\end{abstract}


\section{Preambles} 

The purpose of this contribution is to outline a few key lessons about the nature of time from current physical theories as well as from promising theories under development, and the many ways in which it should be considered an emergent, non-fundamental notion. 
We will be as comprehensive as possible, but still offer only a summary of results and arguments that we have discussed in more detail elsewhere \cite{,Oriti:2013jga,Oriti:2018tym, Oriti:2018dsg,Oriti:2016acw}, while we also develop those thoughts further, in some cases. Also, we will have to leave out all the technical/mathematical material on which our understanding of the issues is based, citing useful references when needed.  
As we will try to convey, there is a lot that can be said about physical time, but it should also be obvious (first of all to readers of this collective book, by a quick look at the table of contents) that \lq time\rq is a multi-faceted notion, so much that one should maybe speak of the many \lq times\rq of our world, from the psychological to the physical, from the physiological to the ecological one\footnote{See \cite{Rovelli:2021elq} and all the resources listed at http://www.studyoftime.org/}. Thus, the first thing to do is to narrow down the scope of our discussion.  

\subsection{Our focus: physical time as deduced from our best theories}
We restrict our attention to physical time only, out of necessity, and of limited personal competence.  

Moreover, we approach the question \lq\lq what is physical time?\rq\rq ~in a rather pragmatic manner (from the perspective of a scientist, at least). We intend by \lq\lq physical time\rq\rq what is expressed in the mathematical models used in physics to account for our experience of the world, i.e. our best current physical theories. 

It is important to notice that the mathematical models used in physics leave out a number of aspects of the physical world and of our interaction with it. Observations and observers are either not modelled at all or highly idealized. That is, many aspects of actual, real world, physical observations are left out altogether (for example details of the measurement apparatuses we use, or physiological aspects of perception). In addition, any physical theory is in itself a model (or collection of models) of a portion of the physical world, and its \lq lessons\rq are thus necessarily partial \cite{sep-models-science}. This implies that our conclusions about the nature of physical time are going to be by necessity the result of some idealization. This is something that we should not see as a problem or a disappointing fact, but something to be embraced, exactly because it leaves much to be explored further, about how to improve the same theories, the relation with other sciences and other aspects of the world, and so on. In any case, we would argue (but not on this occasion), that any understanding is modeling and any model is also the result of some mixture of idealization, abstraction and approximation. 

Our main focus, i.e. to illustrate the nature of physical time as deduced from the best current theories and from those under development, is also the result of a basic naturalistic attitude. Any metaphysics (e.g. any statement about the ontology of time) can only be deduced, or at least strongly influenced by, and should be necessarily consistent with our best scientific theories. We simply state this point here in order to frame our approach better, by being transparent about some underlying assumptions, but it is also one that would deserve to be argued for. Luckily, other competent scholars have done it already, from different perspectives \cite{Ladyman2007-LADETM,Morganti2017-TAHMNM}.

\subsection{Emergence and its many kinds}

The main focus of our discussion of the physical time is its emergent nature. We will argue that this is already visible in classical General Relativity but becomes even more radically manifest when we consider the (possible) implications of a theory of quantum gravity. Thus, it is worth clarifying beforehand what we mean by \lq emergence\rq and to point out the many forms in which emergent behaviour and emergent notions can show up. These different manifestations will also enter our discussion about time in quantum gravity (for early work on this issue, see \cite{Butterfield1999-BUTOTE}). 

In order to develop our arguments, we rest on the notion of emergence put forward by Butterfield and collaborators\cite{Butterfield2011-BUTLID, Butterfield2011-BUTERA} (see \cite{10.7551/mitpress/9780262026215.001.0001} for a broader set of views on the issue of emergence in science): a physical behaviour or phenomenon is understood as {\it emergent} if it is sufficiently novel and robust with respect to some comparison class, usually associated to the class of behaviours and phenomena it emerges from. Very often (but not as a matter of necessity, so to be included in the definition) emergent phenomena are associated to some form of limit or to some approximation\cite{pittphilsci8554, Batterman2011-BATESA, Batterman2004-BATCPA, Castellani2000-CASREA}, applied to the mathematical model describing them and the physical context they emerge from. This definition is simple and general enough to accommodate all known examples of emergent phenomena in physics. It is also somewhat vague, at this stage, but any further refinement would be at risk of restricting too much its scope (leaving out some interesting physical example) and forcing us into unnecessary complications for our purposes. It agrees with the routine use of the term in physics (which is in fact even more vaguely defined). It should be taken as a first step toward a full characterization and understanding of specific emergent phenomena, not the final step, something flexible enough to be adapted and deepened when needed, in different contexts. 

One key feature of this definition is that it is not in contradiction or incompatible with reduction (this being basically understood, within a mathematical modelling of the relevant phenomena or a logical analysis of the corresponding explanatory links, as deduction). Indeed, reduction is usually needed to specify the comparison class for identifying the novel and robust elements of emergent phenomena. The link between emergence and reduction can be so strong that one could often understand one as the converse, complementary process of the other, at the epistemological level. In the context of mathematical modelling of physical systems, one often understand some phenomenon as {\it emergent} from another exactly {\it because} it has been shown to be deducible (i.e. mathematically derivable, within all sorts of approximations) from the other. It goes without saying, then, that we are speaking here only of what is usually referred to as {\it weak emergence}, as opposed to {\it strong emergent behaviour} \cite{Chalmers2006-CHASAW, 10.7551/mitpress/9780262026215.001.0001}, with which we will not be concerned.

Assuming the same naturalistic attitude we stated above, thus not positing any a priori ontology to which then force our scientific theories, but reading it out (with all inevitable ambiguities) from them, emergent behaviour poses an immediate dilemma: which of side of the emergence/reduction relation should be assigned an ontological status? The \lq fundamental\rq one only, to which the emergent phenomena can be reduced? or do both the emergent and the fundamental sides have a correspondent ontology of equal metaphysical weight? In the following, we do not assume a \lq fundamentalist\rq ontology, i.e. we do not assume that the non-symmetric relation between \lq fundamental\rq and \lq emergent\rq phenomena is also a primacy relation at the ontological level, adopting instead an implicit multi-layered ontology\footnote{This is not to say that the \lq fundamentalist\rq view is not reasonable or correct as a metaphysical position, but just that we do not assume it.}. In the case of time, our arguments supporting the idea that \lq time is emergent\rq, do not imply automatically, in our view, that \lq time does not exist\rq {\it tout court}. In case, the latter statement should be argued for. 

Last, we point out several sub-notions of emergence (see \cite{Rueger2000-RUEPED,Yates2013-YATE,Sartenaer2018-SARFE, Bishop2020-BISCEO-3} for more details). All of them enter our discussion about time as an emergent notion. 

The first kind of emergence is {\it inter-theoretic emergence}, basically understood as the converse of inter-theoretic reduction. Here, we speak of a set of physical phenomena as emergent from another, if the theoretical description of the latter can be reduced to the one of  the former (i.e. the theoretical description of the former can be deduced, under some appropriate procedure involving limits, approximations and more) from the theoretical description of the latter. 

The second kind is ontological emergence, when we say that a set of entities is in fact emergent from another, though some physical process), and under the assumption that the relation is not symmetric and definitely not one of equivalence. This second kind is distinct from the first, generally speaking, but, within our naturalistic assumption, it ends up being so strictly associated with it to be indistinguishable for all practical purposes. 

A third kind of emergent physical behaviour is associated to situations, in which one set of phenomena is replaced by a (often radically) different one as one considers different values of some \lq control\rq physical quantities or parameters, but without any real asymmetry in the relation between the two sets. One can speak of this kind of emergence when referring to a specific dynamical process, in which the control physical quantities evolve in time from one value to another, producing the change in associated physical phenomena; but one can also speak of this kind of (symmetric) emergence, as a change in theoretical description, if referring to the mathematical models used to describe both sets of phenomena and depending on the control paramaters. Due to the symmetric nature of the relation, speaking of emergence in this case can be seen as an abuse of language. However, on the one hand, this use of the term is found regularly in the physics literature (and very often in the philosophical one too); on the other hand, our definition of emergence, given above following Butterfield et al did not include any asymmetry condition, thus we have no reason at this stage to exclude tis case from consideration. We point out that no reduction is involved by this third kind of emergence, per se, even if in all examples we can think of there is an underlying reduction relation, not between the two sides of this \lq\lq emergence relation of the third kind\rq\rq, but between each of them and a third set of phenomena, common to both. 

Looking at the concrete physical situations in which these three kinds of emergence are identified, they are often labeled as synchronic emergence, for the first two kinds, and diachronic emergence, for the third. This is because the third is often associated to physical processes taking place {\it in time}, while the first two do not refer to temporal change at all. As we will discuss, this terminology is problematic, first of all because at the theoretical level, also the third kind of emergence can be defined in a way that makes no reference to temporal evolution, but also, and most importantly for our present purposes, any implicit or explicit reference to time would obviously mess up the application of these concepts to the case of time itself and for the understanding of its own emergent nature.  

All these kinds of emergence should be considered in the case of spacetime in quantum gravity, and time in particular, as we do in the following.  Our discussion should be seen in the context of a growing body of work on spacetime emergence in quantum gravity in the philosophy literature \cite{Huggett2021-HUGOON-3,Crowther2020-CROABS-4,Baron2019-BARTCC-15,LeBihan2019-LEBHWL,Huggett2018-HUGTTE,LeBihan2018-LEBSEI-2,Crowther2014-CROAOO}, in addition of course to the quantum gravity literature. 

\subsection{Making things concrete: an example of emergent behaviour} 
The above discussion is certainly overly sketchy and abstract. Therefore, let us give a concrete illustration of the various notions of emergence, using an example that is as uncontroversial as it can get, at least from the point of view of the usual physics parlance\footnote{Still, this example has been extensively analysed and discussed in the philosophical literature, unraveling a number of interesting conceptual subtleties.}. 

Consider the physical system identified as water molecules\footnote{In fact, for what follows, any atomic system would work}. They are well described in the mathematical language of non-relativistic quantum mechanics, with a Schroedinger evolution equation for their quantum states, an Hamiltonian encoding the relevant forces between them and, in case, external ones they may be subject to, and their quantum observables: their individual momenta and positions, their angular momenta etc. This description is in principle valid for any number of them, but when their number is high, collective effects become mostly important, the relevant physical quantities we want to have control of are different, and the quantum description in terms of the individual molecule states is not manageable, and in fact irrelevant for most practical purposes. We usually switch then to a description in terms of (quantum) statistical mechanics, which is the relevant mathematical context also if we are interested in their behaviour at non-zero temperature, which is basically always the case. Other macroscopic and collective quantities, like the pressure they are collectively subject to or the total occupied volume, become important at this level of description. 

While the quantum properties of water molecules and their constituent atoms are crucial for their microscopic physics, as well as for their higher level chemical consequences, their macroscopic properties are not much determined by quantum features, and if they are what we are interested it, we can study the system in a classical approximation. In fact, we know that in a given range of temperature and pressure, and when the number of molecules is sufficiently large, the description of the system as a single entity called water, a fluid, rather than a collection of molecules, and the theoretical framework of hydrodynamics, is the most efficient way of understanding the system. At this level of description, new dynamical quantities like the density of the fluid or its collective velocity, and new physical concepts like viscosity, vorticity, and so on, are the relevant ones. 

The situation is more complex than this, though. At the same macroscopic level of description, for large numbers of molecules, but different ranges of temperature and pressure, what replaces the microscopic description in terms of molecules is not hydrodynamics, but one in terms of lattice structures and their deformations. Liquid water turns into ice, a solid, whose physics is very different from that of liquids and requires new concepts and different mathematical tools. And at yet other values of temperature and pressure, without changing the microscopic description of the system, we have vapour instead, with yet another description and different physics. Liquid water, vapour and ice are different macroscopic phases of the same microscopic system, separated by phase transitions. 

Where can we speak of {\it emergence} in all this? At several places, in fact \cite{Batterman2006-BATHVM,Batterman2013-BATTTO-5}.  

The very step from a quantum to a classical description of the water molecules, before one considers their collective behaviour, is often taken as an example of emergence, that of a classical world from the quantum one. There is no doubt that classical properties are novel with respect to the typical quantum behaviour, and robust enough that we can often forget the underlying quantum world. It is then a case of emergence, but one that does not entail any ontological emergence, since the basic entities at both classical and quantum level are the same molecules, that simply change physical behaviour. It is however an inter-theoretic emergence, since we do indeed change theoretical framework to describe the same molecules in the two regimes, from quantum to classical mechanics. The move from the molecular dynamics to the fluid hydrodynamics is much more radical, though. It is a case of inter-theoretic emergence as well, but one that entails a switch to a different set of dynamical variables and observables, and marked by the appearance of new concepts altogether, simply not applicable before the switch (what is the viscosity of ten molecules?). At the same time, we know that all the above is not incompatible with reduction, since there is a clear sense in which we can reduce hydrodynamics to molecular dynamics and, for example, define hydrodynamic variables in terms of suitable averages of molecular ones. 

The counterpart of this radical inter-theoretic emergence is also a change in the basic ontology, if we base this on the theoretical description of the system: a collection of molecules, first, each with associated particle-like properties, and a fluid, then, described by continuum fields (density and velocity). We have ontological emergence too. This is a consequence of a different type of limit/approximation that the quantum-to-classical one: a continuum and thermodynamic limit, without which we would not have a liquid-like continuum phase. 

The distinction, conceptual and mathematical, between classical and continuum approximation, is a crucial point also in the quantum gravity case. In fact, not only the two limits are very much distinct, but the order in which they are taken has, in general, important consequences. If we had considered helium-4 atoms instead of water molecules, approximating them first with classical entities and then considering their continuum limit would have given a very similar hydrodynamic behaviour as that of water; on the contrary, taking the continuum limit while retailing their quantum properties would have led us, in the appropriate range of temperature and pressure (plus a few other conditions), to superfluid hydrodynamics first, as a macroscopic consequence of their quantum statistics. One needs a further classical approximation at the continuum level to recover standard hydrodynamics. The many striking properties of superfluids would have remained hidden, if we had conflated, conceptually and mathematically, the two types of limits. 

The result of this limit, however, is not unique. Depending on the value of temperature and pressure, here treated as external control parameters of the theoretical description, from the same molecular system we could arrive at a solid system or at vapour: the system can organize itself, macroscopically, in different inequivalent continuum phases, which correspond, in fact, to different macroscopic systems with very different properties (and ontology?) and which can all be said to be emergent with respect to the molecular one. These are all examples of inter-theoretic and ontological emergence\cite{pittphilsci8554, Batterman2011-BATESA, Castellani2000-CASREA}. 

Notice that the relation between the two descriptions, and the corresponding physical systems, is not symmetric. There is a clear sense in which the fluid comes from the molecules and not viceversa (not least, because also ice and vapour come from the same molecules). Notice also that there is no dynamical process involved, i.e. nothing that the molecules \lq do\rq to become liquid water or ice, no dynamical evolution from molecules to water or ice. It is the theoretical description that changes, not the molecules or the fluid. We can speak of synchronic emergence. However, the existence of different continuum phases allows for phase transitions. These are switches between different theoretical descriptions of the same system (from the microscopic point of view) or between two different physical systems with their associated theoretical descriptions (from the macroscopic point of view). But they can also be seen as actual physical processes, taking place (in time) when the control quantities (temperature and pressure) change. Phase transitions can be seen then as examples of emergence of the third, diachronic kind mentioned above. 

Three points are worth remarking, to conclude. First, in the case of phase transitions, the emergence relation is symmetric; there is no sense in which one would say that ice is more fundamental than liquid water or viceversa, either at the ontological or at the theoretical level, and they are on equal footing with respect to the molecular dynamics. Second, while it is a fact of life that the phase transition can be a temporal process (ice melts, and liquid water boils), its usual description is in terms of equilibrium statistical mechanics where time and dynamics play no role. It is considered an approximation to a more realistic (and much more involved) non-equilibrium description in terms of our standard, non-relativistic temporal evolution with respect to an absolute time, but the equilibrium description is not inconsistent in any way. Third, we find straightforward to interpret the phase transition as a physical process because the relevant parameters can indeed be controlled and made to change from the outside, by the observer or the experimenter manipulating the collection of molecules in the lab (or at home). The last two points will be especially relevant in the discussion of the emergence of phase transitions in quantum gravity, since in that context the situation is much more tricky.

\section{What is time, in General Relativity}

Let us now turn to time in General Relativity, our best theory of time, space and geometry and, indeed, gravitational phenomena. The reason for the latter link is that gravitational phenomena, in General Relativity, are equivalent to geometric properties of spacetime, i.e. statements about duration of time intervals, distances between objects, relations between reference frames, i.e. the very notions of space and time directions used by different observers, and so on. Gravity is spacetime geometry, in General Relativity, that is its main lesson. And this encapsulates our current best understanding of both gravity and spacetime. So we take it seriously and ask what is time, in this context. 
\subsection{Manifold points, diffeomorphisms and background independence}
General relativity is a theory of continuum (in fact, smooth) fields defined on a (differentiable) manifold (i.e. a set of points with appropriate regularity properties). Thus, the ontology behind the theory is given by these elements: fields and manifold. Often, we label the manifold a \lq spacetime\rq ~manifold and identify its points as \lq spacetime events\rq, i.e. the {\it loci} where and when things happen. Among the fields, the metric field enters any determination of geometric properties of spacetime and, in a way, of spacetime itself. We routinely speak of spatiotemporal distances between events, spacetime curvature at a specific event, time elapsed between two events taking place at different points in space, and causal relations between events. All these quantities are functions of the metric field, leading to the identification of the metric field, thus the gravitational field, with the geometry of spacetime itself as instantiated by a manifold of events. This is all good, and, as a convenient fiction or an approximate substitute for a more rigorous story, routinely used with success in doing physics. Taken more literally, though, it is unsatisfactory. 

The reason is that General Relativity is invariant under diffeomorphisms of the manifold, mapping the points of the manifold to one another, and the consequent transformation of all fields defined on it, mapping the values they take at different points in the manifold to one another. The values of fields at different points in the manifold are physically equivalent, if they are related by a diffeomorphism transformation\footnote{All this can be expressed in terms of coordinates, as often done, by identifying diffeomorphisms between points as changes of coordinates at the same point; however, this can be very misleading, since physical theories can be written in coordinate independent manner and this does not change in any way their symmetry properties with respect to diffeomorphism transformations.}. 

The debate on the precise implications of this fact is still active \cite{Pooley2013-POOSAR,Hoefer1998-HOEAVR, Curiel2014-CUROTE-4}, but we subscribe to the view (rather predominant in the physics community and in a good part of the philosophy community) according to which this means that the manifold and its points do not really carry any physical meaning. They are thus not part of the ontology of the world, if not as providing global (topological) conditions on the fields. The world is made of fields and fields only. Moreover, diffeomorphism invariance is closely related (in fact, up to some additional subtleties, it can be identified) with background independence\cite{Giulini:2006yg, Gaul:1999ys}. That is, all fields are dynamical entities, subject to their \lq equations of motion\rq constraining their allowed values and mutual relations. And generic solutions of the equations of the theory, i.e. generic allowed configurations of fields, possess no feature that can be used to single out a preferred direction of time or space. 

This absence of a preferred, non-dynamical, absolute notion of time is what is often indicated by the statement that \lq there is no fundamental time\rq in General Relativity \cite{Rovelli:2009ee}, contrary to what is the case in all non-general relativistic physics, including standard quantum mechanics and quantum field theory, and it is a fact. We can get around this uncomfortable fact only when working with specific solutions of the theory, like the flat Minkowski geometry of special relativity or other highly symmetric configurations of the metric\footnote{Or special boundary conditions, corresponding again to highly symmetric geometries.}; in this case, the symmetries of such configurations allow to identify special directions on the manifold as a preferred temporal (or spatial) direction and we deal with an absolute (maybe up to some further transformations, e.g. Lorentz) global time. But beyond these special cases, there is no (preferred) time in GR. 

\subsection{Relational time (and space)}
If what we are left with are dynamical fields, including the metric (i.e. the gravitational field), where is time in GR? and what do we mean by events and their temporal interval? and how is it that we can routinely use coordinates and manifold points to identify events, and functions of the metric defined at such points to measure time? The general answers to these questions, at least as a matter of conceptual clarity and formal mathematical constructions (the detailed physical constructions can be problematic), are given by a \lq relational strategy\rq \cite{Rovelli:1990ph,Marolf:1994nz,Rovelli:2001bz, Dittrich:2005kc}. If the only things that exist are fields, the only physical observables are relations among (values of) fields. In particular, any quantity that we interpret in spatiotemporal manner, in GR, is a relation between specifically chosen fields, used to {\it define} clocks and rods and, by doing so, time and space. All such quantities that correspond to some determination of geometric properties of time and space are then appropriate functions of fields necessarily including the metric. In this sense, also in any rigorous construction of spatiotemporal quantities, the gravitational field will be a necessary ingredient, justifying its often stated identification with spacetime itself. 

Let us give a sketchy example. Instead of having two fields each evaluated at point in \lq\lq time\rq\rq $t$ identified with the value of some coordinate along a given manifold direction on a manifold, we use one of the field as a clock and its values to define instants of time, and measure the evolution of the (values of the) other field as a function of (the values of) such clock (the coordinate time and the direction on the manifold have then entirely disappeared from the physical picture, as they should). The theory itself does not select any physical field as a preferred choice of clock (thus, time). This is the more physical way of characterizing the general covariance of the theory, without referring to unphysical coordinates or manifold directions. 

Most importantly, in general fields behave like good clocks only approximately, only locally (i.e. for some limited range of their values) or in special (and dynamically chosen) configurations. 

To have a good notion of time (or, equivalently, a good clock) the physical system used to define it should interact too strongly with the other fields (including the gravitational field) or with itself, it should not have exceedingly complicated dynamics, and, for the same system to define a globally valid notion of time, further restrictive conditions should be at least approximately true. In particular, in the approximation (or idealized case) in which they do have a vanishing energy-momentum, and thus vanishing effect on the geometry of spacetime, and vanishing self-interaction, and trivial dynamics, then they do behave like simple coordinates labelling manifold points, and can then be used to define time (and space), and the evolution of other fields, forgetting about their own physical nature. In the end, in General Relativity there is no (preferred, external) time, but there are many (an infinity) imperfect, approximate physical clocks, each providing a possible definition of physical time with its own limited applicability. 

In General Relativity, then, time is a specific (set of) approximate relation(s) between continuous physical fields, and from them it inherits its continuous, ordered (and approximate) nature. A last comment is probably useful. An interesting body of work in the classical and quantum gravity literature concerns the \lq deparametrization\rq of General Relativity in terms of suitable matter fields \cite{Husain:2011tk, Thiemann:2004wk,Giesel:2012rb,Giesel:2017roz}. This entails rewriting the full theory in relational terms with respect to the reference frame defined by them, in such a way that one does not need to deal with diffeomorphism symmetry anymore, that all resulting relational observables are physical and the dynamics takes a more standard form with respect to the time defined by the appropriate component of the matter fields. The drawback of this type of strategy, at the classical level, is that the matter fields that allow for this type of rewriting are always somewhat unphysical. In turn this is basically inevitable since they have to provide exactly that type of global, perfect clock and rods, with associated globally defined notions of time and space that, as we discussed, we expect being no more than an idealization or the result of some approximation.  

The above means that, from the point of view of General Relativity, time as an absolute, uniquely identified notion, like in Newtonian mechanics and pre-relativistic physics, is an {\it emergent} notion. It appears for special configurations of the gravitational field, in the approximation in which all relativistic effects are suppressed and all physical clocks (that is, other matter fields used as such) behave uniformly, in addition of being assumed to have no impact on other dynamical entities (so that they can be treated as  measuring an external parameter). It is then an instance of inter-theoretic (or vertical) emergence. Whether it can also be considered as an example of ontological emergence is more dubious. While in the step from relativistic physics to newtonian mechanics fields are replaced by particles and forces as fundamental entities (this is indeed an example of ontological emergence), those fields whose relations define time in GR would stop being part of the dynamical entities of the world and give rise to an absolute, external, non-dynamical notion of time, that can hardly be considered a physical entity on its own.

\section{What is time in quantum gravity, if this is quantized GR}
We expect, however, that General Relativity is not the final story, and that there is more to say, about time in physics. We expect this, for example, because of puzzling aspects of astrophysics and cosmology like dark matter and dark energy, that may be explained by modified gravity theories at the classical level. Most importantly, we expect that General Relativity or other classical modified gravity theories should be replaced by a quantum theory of spacetime, geometry and gravity at the more fundamental level. The arguments for this conclusion are many and, even if not fully conclusive, make the conclusion rather consensual in the community. So we should consider what happens to time in quantum gravity. The detailed answer will depend on the specifics of the 
theory, but such more fundamental theory of quantum gravity has not been established and there are several candidates, which are in fact quite different in their basic mathematical structures and principles, although they also share many ingredients. We have to content ourselves, then, with a less detailed answer based on general aspects shared by several quantum gravity formalisms or simply part of the definition of the quantum gravity problem.

\subsection{Quantum GR, quantum fields, quantum (relational) time}
We would identify as a theory of quantum gravity and quantum spacetime, roughly speaking, any quantum theory that reduces to (some modified version of) General Relativity in a classical (and probably macroscopic) approximation. We should consider also the possibility that the quantum formalism itself has to be modified to be applied to spacetime, but let's assume for now that this is not the case, for simplicity. Then, the simplest possibility, at least conceptually, is that the fundamental quantum theory is the result of \lq quantizing the classical gravity theory\rq by one of the many quantization algorithms we have successfully applied to other field theories (canonical, path integral, etc). Several approaches to quantum gravity can be understood from this perspective \cite{Kiefer:2013jqa,Kiefer:2012cdl, Thiemann:2007pyv}.
 
 Then, the ontology of the resulting quantum theory is the same as that of the classical one. The world remains constituted by continuum fields, among which the gravitational one is the one strictly associated to geometric spacetime properties. Physical notions of space and time, as in GR, remain dependant on relational constructions involving several (components of) dynamical fields, except in very special cases where preferred temporal or spatial directions can be identified. However,  the same relational constructions should be performed at the quantum level, and the same fields become quantum systems, starting from the gravitational field. This step to the quantum domain brings radical changes for our notions of space and time.

The quantum nature of the gravitational field implies the quantum nature of spacetime geometry and the quantum nature of all the other dynamical fields implies the quantum nature of any notion of time constructed using them as relational clocks (more generally, reference frames). Possessing a quantum nature means being subject to uncertainty relations, irreducible quantum fluctuations, some form of contextuality, discreteness of observable values, and, in the case of composite systems, entanglement, in turn challenging our common sense notions of realism, separability, and locality. When these properties are attributed to geometric and spatiotemporal quantities, we clearly enter a new and wild conceptual (and physical) domain. All geometric quantities, like areas of surfaces, distances and temporal intervals between events, curvature in a region, are then subject to superposition and quantum fluctuations, they may be forced to have only discrete values, they may be incompatible with one another (like position and momentum of a quantum particle), and they may be restricted to a contextual-only specification. The causal structure of spacetime itself (i.e. the list of potential cause-effect relations) will be similarly subject to quantum fluctuations and superpositions. And the list of quantum weirdnesses of a quantum spacetime could go on. The relational strategy for the definition of time will be further affected by the quantum properties of any physical field we choose as our relational frame, in particular our clock \cite{Hoehn:2019fsy,Castro-Ruiz:2019nnl,Giacomini:2021gei}, itself subject to quantum fluctuations, uncertainty relations etc. 

So time will be whatever it was in GR, but quantum. But a quantum time is even farther away from any standard notion of time on which our common sense, and our classical physics is based, that we can definitely say we are in a very different world, or better at a very different, more fundamental level of our understanding of this world. The physical and philosophical literature has barely started to explore these new conceptual depths, also due to the limitations of our current quantum gravity formalisms, but it should be clear that we will  need a profound reconstruction of the basic pillars of both our physics and our philosophy to account for these new aspects. With this quantum step, we must abandon, for example, any value-definiteness of spatiotemporal quantities and possibly any continuous notion of space and time, if spatiotemporal observables end up having discrete values \cite{Rovelli:1994ge}. The idea of quantum reference frames and indefinite causal structures or temporal order, moreover, represent new challenges for the foundations of quantum mechanics themselves \cite{Castro-Ruiz:2019nnl,Giacomini:2021gei}, and together with the notion of a preferred time direction we are forced to abandon also unitary time evolution as the key dynamical element of quantum mechanics. To make sense of dynamics in absence of such unitary evolution is sometimes referred to as \lq the problem of time\rq in quantum gravity, and it is the direct quantum counterpart of the diffeomorphism invariance of classical General Relativity (as such, it affects space in the same manner). As we argued, there is no \lq solution\rq to this problem, strictly speaking, as long as the symmetries of the classical theory are preserved \cite{Isham:1992ms,Bojowald:2010qw,Marolf:2009wp}. The absence of a preferred temporal direction is a fact, and the relational strategy is the best way, we maintain, to extract from the theory a physical, if approximate, notion of time and evolution. 

\subsection{The timeless emergence of time from a quantum time}
The quantum aspects of the story are not minor additions and they fully justify to speak of the emergence of the general relativistic time from quantum gravity. To start with, they prevent any of the circumvention strategies of the problems given by the absence of a preferred time direction that are  available in classical GR, e.g. working with special solutions possessing preferred temporal directions (in the quantum theory there is no state that corresponds to a classical solution of the theory, if not approximately only). This is the sense in which the \lq problem of time\rq is worse in quantum GR than in classical GR, even if its origin is the same, i.e. diffeomorphism invariance. 

The set of approximations and mathematical procedures, choices of special states and focus on specific observables that lead to classical GR from quantum gravity (realized to various degrees of success in current quantum gravity formalisms, see for example \cite{Alesci:2014uha,Barrett:2010ex,Asante:2021zzh} in the loop quantum gravity and spin foam context) that go under the collective label of \lq classical limit\rq  would also lead to recovering the temporal observables computed as relations between classical fields, from corresponding quantum observables. There will be novelty enough, since the classical limit brings continuous quantities where, probably, there were discrete ones, value-definiteness where there were indefinite temporal order and fluctuating temporal durations, and the classical dynamical aspects of time and its relation to space where we had the limitations of contextuality and non-commutativity of quantum observables. And certainly the classical world described in its spatiotemporal aspects by General Relativity is robust enough, since quantum features of gravity and spacetime are so hard to detect (so much that we have very little guidance from observations, in our search for the more fundamental quantum gravity description). In this respect, we can see also the limitation of the global deparametrization strategy at the quantum level. Deparametrizing the theory at the classical level and then quantizing , e.g. by {\it reduced phase space quantization} \cite{Thiemann:2004wk,Giesel:2012rb,Giesel:2017roz} in terms of the canonical decomposition defined by the matter fields introduced as clock and rods, requires neglecting the quantum nature of the clock field itself, thus of an important aspect of its physical nature, and of the effects that its quantum properties may have on other fields.. Again, we see that a global notion of time, convenient as it may be and also in this more physical sense of a relational clock, remains an idealization or something that can be valid only in some special approximation of the fundamental theory. 

This further sense in which time as we know it is emergent, this additional level of emergence is again, first of all, inter-theoretic and synchronic; it corresponds to a change of theoretical framework and associated conceptual one, but it is as such not a physical process in itself and by definition not a temporal one in any case, thus it would not qualify as an instance of diachronic emergence. It is not an ontological emergence either, since we have assumed that the fundamental entities in quantum gravity are the same fields that GR deals with, only turned into quantum entities\footnote{In our analogy with water molecules and liquid water, we remain at the hydrodynamic level with the system of interest being liquid water and we move to a regime in which the quantum nature of the fluid (not the constituent molecules, whose existence we simply ignore and never enter the picture) can be neglected altogether. The analogy is not so compelling in this case, since liquid water is a fluid whose macroscopic quantum properties can basically always be neglected. A better analogy would be with superlfuid, which manifest macroscopic quantum behaviour and thus a non-trivial classical limit.}.

We lack a complete theory of quantum gravity, though, so we do not know the precise details of the physical circumstances that allow us to pass from the fundamental description of the world, in which all fields including the gravitational one, and  thus spacetime, are quantum entities, to the one of GR, in which spacetime, geometry and all dynamical fields behave classically. In this situation, it is hard to characterize clearly how this transition to a classical world comes about. We expect it to be the result of a physical process, not only a shift in the most appropriate theoretical description. It should take place whenever the relevant length scales, curvature scales and related change value from trans-Planckian to sub-Planckian ones\footnote{Let us stress that this expectation is in line with our effective field theory intuition and based on established general relativistic and quantum physics, therefore very solid but also subject to change in a more fundamental theory.}. This has happened, for example, in the very early universe, close to the big bang singularity predicted by the classical theory. In fact, many quantum gravity theorists and cosmologist would expect just that: a transition from a fully quantum epoch in which full quantum gravity is the correct description of the world, close to the big bang, to a classical one governed by General Relativity (better, semiclassical gravity with quantum matter fields living on classical spacetime), when the universe is larger and it has \lq classicalized\rq in its spatiotemporal aspects, due also to the interactions between quantum matter fields and quantum geometry. 

This is all good, as far as the general idea of a quantum-to-classical transition being a physical process is concerned. It becomes problematic if we take seriously also our temporal language and interpret the transition as something happening {\it in time}. 

Consider the picture in more detail, as we move backward towards the early universe, starting from our present classical one. We select a time variable, first of all, to be allowed to speak of early and late universe and of cosmological evolution. The specific choice is not so relevant, now, but it corresponds to some physical degrees of freedom (a field, a component of it, or a function of it) being used as a clock, with all the other physical quantities expressed as a function of its value. This value should be well-defined, to be used as a good reference. In a quantum world, this means that it corresponds to a semiclassical observable, whose mean value we use as reference value and whose quantum fluctuations are negligible compared to it (otherwise we would not be able to \lq follow\rq the evolution at all\footnote{Notice that the possibility that we use instead the eigenvalues of an observable chosen as clock is problematic, although often studied \cite{Agullo:2016tjh}, because in this case we would be considering states in which the conjugate observable to the clock time is maximally uncertain, and we should expect these large quantum fluctuations to be physically relevant to, impacting somehow on the evolution of the other physical systems. We would be dealing in fact with a highly quantum state of the system; it is not the kind of time variables we use in cosmology.}. Then we follow the evolution of the universe and all that it contains in relation to this clock, toward earlier epochs. At some point (in clock time) quantum aspects of the world (i.e. of all the physical systems) start becoming relevant. As we move further toward the big bang, they become even more relevant and we enter the full quantum gravity domain. The transition is then understood as a physical and temporal process. 

Not so fast, though. For this to be the case, the temporal observable defined by our clock should remain (approximately) classical, otherwise the very notion of relational time evolution would cease to be meaningful. Even within a full theory of quantum gravity, there may well be a physical situation like this, with a specific subsystem, which happens to be the one we have chosen as a clock (maybe exactly for this reason) remains classical also while all the others start manifesting strong quantum fluctuations and other quantum properties. But then there is no emergence of classical, relativistic time from the quantum version of the same we have in the full quantum gravity theory, in such transition. It does not correspond to the situation we wanted to interpret as a physical process also with respect to time itself. In a situation in which, on the contrary, as we follow the evolution of the universe backwards in relational time, our clock itself starts manifesting more and more its quantum nature, we do indeed face the type of transition from quantum time to classical time (or vice versa, in this case)  that we were considering to speak of emergence of classical time as a dynamical, physical process, but its own temporal characterization stops exactly where (or when) the transition occurs. Beyond that, in the full quantum regime in which we only have quantum time (and space, and geometry, and causality) any standard notion of temporal evolution (including the relational one) stops being applicable. No diachronic emergence, again. The emergence of time remains timeless.

We are going to encounter again this type of situation (an inter-theoretic emergence that is not ontological, that could be in correspondence with a physical process, that can be \lq met at the end point of some temporal evolution\rq ,  but that remains timeless) again in the following.

\section{What is time in quantum gravity, if this is {\it not} quantized GR}
The story of the progressive disappearance of time in a quantum gravity context, when moving toward the more fundamental level, and of its emergent nature when seen in the opposite direction, takes a more radical turn in quantum gravity formalisms in which the theory is not the straightforward quantum version of classical General Relativity (or other gravitational theory for the metric field coupled to matter fields). By definition, these formalisms are defined in terms of a different ontology. The fundamental entities are not continuum fields (including the metric), be them classical or quantum, and such continuum fields are themselves collective, emergent entities. 

\subsection{A timeless ontological emergence of time}
What the new fundamental entities are depends on the specific formalism one considers, and similarly specific is how different they are from the usual spatiotemporal ontology of fields. The notions of space and time as encoded in field-based observables in General Relativity are complex notions, we we have explained, involving topological and metric (extension) features, continuity and directionality, causal and relational aspects. Therefore, we are in presence of new entities, whenever we do not deal with continuum fields, and with a more radical absence of the usual notions of space and time, i.e. a non-spatiotemporal level of understanding the physical world, whenever one or more fo these features are absent or modified. And the more ingredients of the usual complex notions we drop, the farther away from space and time we end up being. 
We stress that this is an independent step away from spacetime as we know it with respect to the upgrading of the classical constructions to the quantum level, distinct from the inclusion of the quantum properties of fields into the picture. 

Examples of candidate fundamental entities of non-spatiotemporal nature, in the above sense, are the spin networks and spin foams in canonical Loop Quantum Gravity \cite{Sahlmann:2010zf}, spin foams models \cite{Perez:2012wv} and group field theories \cite{Oriti:2014uga}, the abstract simplices and associated piecewise-flat geometries in lattice quantum gravity \cite{Loll:2019rdj}, spin foam models \cite{Perez:2012wv}, and tensor models and tensorial group field theories \cite{Gurau:2012hl,Oriti:2011jm,Krajewski:2012aw}, the posets of causal set theory \cite{Surya:2019ndm}. Also in string theory the same idea of an emergent spacetime is central, thanks to non-perturbative results like the string dualities \cite{Blau:1900zza}, even though there are less clear suggestions for what the fundamental non-spatiotemporal entities could be; in proposals like the IKKT Matrix Theory \cite{Aoki:1998vn}, for example, they are in fact of a similar combinatorial and algebraic nature than in the other approaches.  

This is a very radical step. The big challenge at the conceptual level, beyond the physical and mathematical ones, is to gain any sort of intuition of the nature of the new fundamental entities, since our physical thinking is grounded solidly in space and time. And understanding is more than just intuition. This change in fundamental ontology requies by definition the development of a metaphysics in which the usual notions of space and time, thus location and ontological distinction based on it, spatial contiguity, temporal change and permanence, and so on, do not feature in the very definition of existence or reality. It is quite a challenge for contemporary and future philosophers, and an important one. 

A second set of ontological worries concerns instead spacetime in this emergence scenario, and it has to do with how the ontological status of the entities featuring in our physical theories is affected by established emergence relations.  If one adopts a \lq fundamentalist ontology\rq according to which only the entities featuring \lq fundamental\rq theories are entitles to ontological status, while those appearing in emergent or effective descriptions do not, the logical conclusion is that space and time cannot be real entities. This metaphysical position is questionable in itself, even before considering quantum gravity. Think of the paradigmatic example of emergence given by molecules and fluids, where this position would imply that fluids, actually, do not exist in a proper sense. It is clear, however, that a new level of philosophical difficulties arise when it is space and time that are deprived of their status as elements of reality. The way out is some form of \lq multi-level ontology\rq in which both fundamental and emergent entities are understood as real. How articulate in detail and in a solid manner this form of ontology, however, is itself an interesting and non-trivial challenge.

\subsection{The timeless emergence of time via a continuum approximation}
Let's move back to the physical aspects of spacetime emergence in this scenario. Starting from the new candidates for fundamental quantum entities of the world proposed by one or the other of the current quantum gravity formalisms, the emergence of spacetime corresponds then, first of all, to the emergence of fields, including the metric, and (a modified version of) General Relativity from their collective quantum dynamics. Once this step is taken, the usual (relational) spacetime constructions and notions of relativistic physics can be used, and, further down the line, the common sense notions of space and time can be obtained in correspondence with special cases or further approximations. This step is what goes under the generic label of \lq continuum approximation\rq in quantum gravity approaches, since in most cases the new non-spatiotemporal entities are discrete in nature, in one form or another. To understand how to perform this step is (one of) the main outstanding issue in all quantum gravity approaches. To do so, these approaches have to adapt to a background independent context, first, and then apply the same methods routinely used in quantum many-body theory to extract macroscopic and collective physics from the fundamental quantum dynamics, e.g. coarse graining techniques and renormalization group. This is if fact a very active research area in fundamental quantum gravity approaches \cite{Bahr:2017klw,Dittrich:2014ala,Eichhorn:2018phj,Carrozza:2016vsq,Finocchiaro:2020fhl}.

The result would be to pass from the \lq atomic\rq description of quantum gravity to the quantum gravity counterpart of an \lq hydrodynamic\rq approximation in terms of continuum quantities, within which one could then extract a classical General Relativistic dynamics for spacetime and geometry. In some approaches to quantum gravity, like tensorial group field theory, this hydrodynamic analogy is in fact quite literally realized, in order to extract some effective gravitational (cosmological) physics \cite{Oriti:2016acw, Pithis:2019tvp, Gielen:2016dss}. The extraction of gravitational dynamics and its approximate rewriting in terms of temporal evolution would then be performed at the effective, hydrodynamic level, adopting once more the relational strategy in terms of the emergent fields (for an example of this procedure, in the same quantum gravity formalism, see \cite{Oriti:2016qtz, Marchetti:2020umh}).

If the starting point is quantum and the result is classical, a \lq classical approximation\rq should be involved as well, together with the continuum approximation. However, we stress that this is a distinct step, conceptually, physically and mathematically. Depending on the details of the quantum gravity formalism and of different physical situations, it may take place in conjunction with the continuum approximation, having them somehow intertwined. However, being distinct, they may take place independently. In this case, one should further notice that the order in which they are taken is not irrelevant. Experience with quantum many-body systems teaches us that the quantum properties of the fundamental constituents may be crucial for capturing the correct macroscopic properties of the collective system formed by them. In the analogy with atoms or molecules and fluids, we have quantum atoms whose fundamental description is expressed by a many-body quantum Schroedinger equation or a corresponding quantum field theory, and whose collective physics can be described by the hydrodynamics of a continuum fluid, capturing all the relevant observables and dynamics at a macroscopic scale in which the individual dynamics of the atoms is not of interest and the atoms themselves are not relevant anymore as physical entities. 

But we know very well that there are fluids and superfluids, differing at macroscopic level in a physically (and technological) very relevant manner, and that specific features of the latter are directly due to the quantum properties of the fundamental atoms, e.g. their bosonic statistics. In order to correctly capture these properties one cannot take first a classical approximation of the quantum theory of atoms (after which they would be described, say, as classical particles) and then look for a continuum, hydrodynamic approximation. That is, for many systems it works fine, but for many others it fails to be physically correct, failing to account for macroscopic quantum effects like superfluidity, superconductivity, quantum phases of matter, etc. 

Notice that we are not referring only to the quantum properties of the atomic constituents being visible if one looks carefully enough beyond the continuum, hydrodynamic approximation, or at small enough distance scales or at high enough energy, This is not under question. It is a logical necessity that one can go beyond the continuum description and be able to distinguish physical effects due to the underlying atomic structure. Otherwise the atomic description would not qualify as useful as a physical theory. In the quantum gravity case, there is no question that fundamental quantum gravity models should predict modifications of classical GR at high energies or small distances or due to the quantum nature of fields. The same is true also for quantum gravity formalisms that are the straightforward quantization of the classical theory. 

In a context in which spacetime is emergent in this more radical sense, an entirely different type of quantum gravity effects is at least conceivable. Some of them have to do with the very existence of different fundamental entities, and maybe with their discrete nature, but not directly with their quantum properties. Others will be the result of their quantum properties. In both cases, their collective effects can well be macroscopic, i.e. visible at low energy and large distance scales, since they are not captured by the effective field theory intuition based on given spacetime structures. An examples of physical phenomena that have been discussed in this spirit, in a quantum gravity context, is dark energy \cite{Oriti:2021rvm}. What is the case for spacetime and gravity? Are some of their features directly due to the {\it quantum} properties of the fundamental, non-spatiotemporal quantum gravity constituents? Is our universe akin to a quantum fluid? We do not know yet. We are trying to find out.

\subsection{What kind of emergence is this?}
In terms of classifications, this more radical type of emergence of spacetime in quantum gravity theories is certainly associated to ontological emergence, as we have stressed. And it is, necessarily so in our \lq naturalistic framework\rq an inter-theoretic emergence too. The step from the description of the world in terms quantum gravity atoms to the one in terms of continuum fields may be also made necessary in correspondence with physical situations, e.g. with cosmological phenomena close to the big bang being the natural candidates.  It may well be the case that at the high densities found close to the big bang, at the very origin of our universe, the \lq hydrodynamic\rq description of the universe corresponding to General Relativity breaks down not (only) because quantum effects manifesting the quantum nature of fields become important, but because the very description in terms of continuum fields, thus continuum space and time, break down. This breakdown could then be understood in terms of the properties and dynamics of the quantum gravity atoms themselves, providing then an explanation and a quantitative account for it. The new atomic description offered by a more fundamental theory of quantum gravity would be shown to be necessary. So the emergence of spacetime from non-spatiotemporal atoms, thus also in this more radical sense, would be as physical as as it can get. 

But would it be also an example of \lq diachronic\rq emergence? would it be a \lq temporal\rq process itself? It would be hard to respond in the affirmative, in general. Consider again the analogy with atoms and fluids. There are many physical situations in which the atomic structure of fluids is immediately relevant for understanding their properties, and in which, correspondingly, we have to switch from the hydrodynamic description to the quantum atomic one. The example of superfluidity is again a good one. But we would not say that something has happened to the atoms that has \lq produced\rq the fluid as a result, when we look at a Bose condensate manifesting its superfluidity properties more and more (or less and less) in changing physical conditions. We remain in the context of an inter-theoretic relation, with a solid physical motivation and impact. There may be cases in which a \lq temporal\rq characterization is instead quite appropriate. In the same example of fluids, consider the case in which the fluid becomes less and less dense as a result of its own dynamics.  Then (quantum) statistical fluctuations due to its atomic structure could become, in relative terms, more and more relevant and thus force us to switch to the atomic description of the system. One could then say that the emergence of the fluid from atoms, and vice versa, is associated to a dynamical process as well. 

Bringing this intuition to the case of time in quantum gravity, however, we face exactly the same type of conceptual and technical difficulties we encountered when discussing emergence of (classical) time in quantum General Relativity from the disappearance of quantum effects from the dynamics of fields. We can hypothetically follow the evolution of the universe toward the would-be Big Bang, only to find out that, sufficiently close to it, the whole continuum description of the universe in terms of spatiotemporal fields needs to be replaced by one in terms of non-spatiotemporal quantum gravity atoms, but there is no way we can speak of temporal evolution, once we (have been forced to) adopt such description. 

The only alternative possibility to this conclusion is if some (relational) notion of time is somehow preserved from dissolution into the atomic description, so that it can still be used to describe the dynamics of the \lq atoms of space\rq themselves. For example, this would happen in a formulation of the fundamental theory in which a deparametrized dynamics with respect to some internal clock variable can be adopted {\it before quantization}. with the consequence that the corresponding relational time is treated as an external parameter and can be used to label the evolution of all the other non-spatiotemporal entities. This procedure is used in some formulations of tensorial group field theory, and in its cosmological applications \cite{Wilson-Ewing:2018mrp,Gielen:2019kae,Gielen:2021dlk}, for example. 

However, as it is the case in classical GR, the resulting theory can only be, in general, a special sector of the full theory, whose predictions are at best only approximately valid and only in special circumstances. Specifically, they will only be valid provided one can neglect the quantum and dynamical nature of the degrees of freedom chosen as relational clock. In a more general situation, though, we cannot expect this to be valid. The temporal description of the process leading to the disappearance of time, that is, ceases to be valid exactly when such disappearance takes place, by definition. There is no diachronic emergence across the two descriptions. There can be, at best, a diachronic (i.e. in terms of temporal change) approach {\it toward the disappearance of time}, and a diachronic account {\it from the emergence of time onward}. This is where the analogy with ordinary atoms and fluids, that live {\it in spacetime}, fails and we are left in search of new intuitions\footnote{An important remark should be added. In the context of a quantum gravity formalism that is based on new types of fundamental degrees of freedom, different from continuum fields, one has always the option to take a perspective in which these new structures are purely mathematical artefacts or useful technical tools to arrive at a true physical definition fo the theory that is in fact based on the usual continuum fields. Indeed, several practitioners view in this way some some of the quantum gravity formalisms we have mentioned. In this case, most of the technical challenges in recovering such continuum description and a viable gravitational description would remain exactly the same (they are \lq interpretation-independent\rq). What changes is of course that one would not have to worry about conceptual or physical issues related to the nature of the new fundamental entities (since they would not be \lq real\rq) nor about the nature of a new emergence process, since we would be in the same situation as in the previous section. The other difference is that one now has to impose on the theory the requirement that no sign of the structures used as technical tools to recover continuum physics should survive the reconstruction procedure.They should entirely disappear from the final result in the computation of any physical quantity. This becomes a key constraint in the definition of the \lq continuum limit\rq. If one adopts the opposite viewpoint and regards the new structures as somehow physical or \lq real\rq, then the conceptual issues cannot be avoided. Moreover, some signature of their existence should in fact survive the continuum limit, otherwise it would be entirely vacuous to consider them real. For the same reason, the choice between these two perspectives is not arbitrary but, as always in physical theories, will have to be decided by observations. Only if such observables consequences of the existence of the new fundamental entities can be theoretically derived, first, and then experimentally confirmed, then the \lq realist\rq perspective will find a strong support.}.

\section{Things get worse, for time: quantum gravity phase transitions and geometrogenesis} 
The story of the emergence of time (and space) is even more complex than this. The same is true, in fact, also in the more familiar case of atoms and fluids. 

\subsection{Inequivalent continuum phases and phase transitions}
The point is that the continuum limit, encoding the collective behaviour of atoms and, more generally, quantum many-body systems is not unique: starting from the same microscopic quantum constituents, their collective dynamics may lead to very different macroscopic phases, that is very different macroscopic physical systems\footnote{To tell the whole story, the converse is also true. There are usually different microscopic systems that can give rise to the same macroscopic behaviour, thus the same macroscopic physical system, with their microscopic differences becoming irrelevant in the limit. This is the phenomenon of {\it universality} in statistical and quantum many-body physics. In other words, just like emergence is not a unique relation, reduction is not unique either, in general.}. The same water molecules that give us liquid water in a certain range of temperature and pressure, as their macroscopic counterpart, can also give us vapour when temperatures are higher, and solid ice when they are lower. And thinking again at superfluids, the same Helium-4 atoms that constitute them at very low temperatures will instead produce standard fluids at higher temperatures and gaseous systems at even higher ones. 

The appearance of different macroscopic phases in the continuum limit of quantum many-body systems is the rule, and a constant source of marvel and challenges, concerned with the investigation of the rich set of new features that is shown in the different phases and of the conditions for the phase transitions leading from one to the other. 

What should we expect in quantum gravity, when studying in detail the collective behaviour and continuum limit of the candidate \lq atoms of space\rq suggested by any given quantum gravity formalism? Just the same non-uniqueness of the result of the limit, the same richness of emergent macroscopic physics, the same variety of possible macroscopic phases. Examples of different continuum phases and analyses of the related phase transitions can be found in much of the recent quantum gravity literature, e.g. \cite{Delcamp:2016dqo, Kegeles:2017ems, Bahr:2015bra, Koslowski:2011vn, Surya:2011du, Pithis:2020kio,Ambjorn:2016mnn}. 

If any of the proposed quantum gravity formalisms has a chance of being a viable description of the physical world, and of providing a satisfactory explanation for the emergence of spacetime and gravitational physics from a more fundamental non-spatiotemporal reality, then at least one of the continuum phases it produces should be properly spatiotemporal and geometric. That is, it should allow to reconstruct an effective (and approximate only) gravitational dynamics based on fields, spacetime and geometry and described by (some possibly modified form of) General Relativity. In other words, in at least one such phase one should find herself in the situation described in the previous section and, from there, move to the notions of space and time encoded in usual relativistic physics. This may or may not involve, as an intermediate step, a further approximate regime governed by some form of \lq quantum GR\rq , depending on what is the exact relation between continuum and classical approximations in the quantum gravity formalism being considered. 

\subsection{The fundamental ontology is even more timeless}
If one such phase is a physical necessity for the viability fo the theory, and the existence of several geometric phases (e.g. describing spatiotemporal and gravitational physics around different backgrounds, or governed by different gravitational theories) is also a possibility, the possibility of non-geometric, non-spatiotemporal phases raises many new issues, both physical and philosophical. 

The first implication is that the non-spatiotemporal nature of the fundamental entities is more profoundly so that it could have been suggested if there were only geometric phases in the continuum limit of the theory. Indeed, if the continuum limit were to always produce spacetime and gravitational physics, one could argue that the fundamental entities, albeit not spatio-temporal or geometric in the usual sense, possessed features that would {\it necessarily} lead to the emergence of spacetime under collective quantum evolution; they could be seen as possessing some sort of \lq proto-spatiotemporal\rq ~features, some more primitive form of spacetime characterization, whose relation to spacetime itself was strong enough to to ensure the appearance of the latter after some approximation. One could even argue that, in this case, the usual notions of space and time would need to be adapted in the more fundamental theory, but do not cease to be meaningful, and question the very notion of spacetime emergence or its relevance. Not so if non-geometric phases can be produced by the same fundamental quantum gravity entities. This very possibility eliminates any necessary link between their features (quantum observables) and spacetime: the same quantities that, under some specific conditions, can be used to define and reconstruct time in one continuum phase of the theory, can produce non-spatiotemporal continuum physics under different circumstances (e.g. different values of the parameters entering the definition of the fundamental quantum dynamics). The issue is physical, of course, because it means that the step from quantum gravity to emergent spacetime physics is more subtle and more technically challenging than if it was otherwise, and that the emergent continuum physics is richer. It also implies that the philosophical challenges to be faced when establishing the new \lq quantum gravity ontology\rq of the new non-spatiotemporal entities have to be solved without even the indirect support of spatiotemporal intuitions that a necessary link between them and spacetime would have allowed. In other words, the existence of different continuum phases including non-spatiotemporal ones makes the ontological emergence of spacetime even more radical, emphasizing the novelty of space and time from the point of view of the fundamental quantum gravity constituents of the universe. 

\subsection{Geometrogenesis}
Beyond the ontology of the fundamental entities and the added complexity of the continuum limit that should lead to spacetime emergence, the existence of different phases of the quantum gravity dynamics raises interesting new issues {\it at the continuum level}, related to the physics and philosophy of the phase transitions leading from non-geometric to spatiotemporal phases. Such phase transitions have been dubbed {\it geometrogenesis} in the quantum gravity literature \cite{Konopka:2006hu,Oriti:2007qd,Mandrysz:2018sle, Smolin:2003qf}, and (whether interpreted \lq realistically\rq ~or not) are actively investigated in several quantum gravity approaches \cite{Delcamp:2016dqo, Kegeles:2017ems, Bahr:2015bra, Koslowski:2011vn, Surya:2011du, Pithis:2020kio,Ambjorn:2016mnn}.

A first question is how to characterize the existence of the transition. What are the quantum gravity observables that show the relevant discontinuities\footnote{A phase transition is defined by non-analytic behaviour of some observable} at the transition point? and what are the relevant quantities that can play the role of order parameters, identifying the different phases with their different values? The answers can only be specific to different quantum gravity formalisms, but since it is spacetime and geometry that emerge at the phase transition, it must be geometric and spatiotemporal quantities that characterize it, by acquiring physically meaningful values only in the geometric phase. Various examples have been suggested in the quantum gravity literature, for example the metric field components themselves \cite{Percacci:1990wy} or the universe volume, that would be identically vanishing (in expectation value) in the non-geometric phase and non-vanishing and with a non-trivial dynamics in the geometric one\footnote{This would be in analogy with the Higgs mechanism of spontaneous symmetry breaking, in which the Higgs field acquires a non-vanishing expectation value, giving mass to all other fields, in the broken phase.}.

A second question is of course what are the observable features of such phase transitions. What is the physics of geometrogenesis? To answer this, quantum gravity models should identify geometrogenesis with some precise physical circumstances, and again the most natural suggestion is the physical regime associated, in the classical theory, to spacetime singularities, like the interior of black holes and the cosmological beginning, the Big Bang. The possible answers to this question are also context-specific, with different quantum gravity formalisms offering different proposals. In general, the approach to phase transitions is characterized in terms of strong fluctuations of some key physical quantity (e.g. of the order parameters themselves), thus this second question can only be tackled together with the first one. But the physics of phase transitions in quantum many-body systems is very rich, so the range of possibilities is large and should make us optimistic about the potential testability of a geometrogenesis scenario in quantum gravity. 

It should be clear that the main objective of any phenomenological analysis of the physics of quantum gravity phase transitions should focus on the testable phenomena {\it on the spatiotemporal side of geometrogenesis}, i.e. the geometric phase we live in. This is simply because it is hard to imagine how we could access directly the hypothetical non-geometric phase, of whose existence we should then identify instead some interesting indirect testable consequence. For example, in the scenario in which the geometrogenesis phase transition is how quantum gravity replaces the big bang singularity in the very early universe, one should look for its possible imprints in the consequent modification of the universe dynamics and in the physics of cosmological perturbations in the very first instants of the history fo the universe, in particular in the CMB spectrum. Indeed, some quantum gravity formalisms are investigating this kind of cosmological scenarios, two examples being group field theory condensate cosmology \cite{Oriti:2016acw, Pithis:2019tvp, Gielen:2016dss} and string gas cosmology \cite{Brandenberger:2015kga} (which is\footnote{If the pre-transition phase is associated with vanishing, rather than just constant, scale factor.} strictly speaking an example of the emergence of space only, not time), and some more phenomenological proposals for possible signatures of such phase transition have been put forward \cite{Magueijo:2006fu,Mielczarek:2014aka}. 

\subsection{Geometrogenesis as a physical process: a temporal characterization?}
In the above discussion we have assumed the perspective in which the geometrogenesis phase transition is a physical phenomenon (not just a mathematical artefact), and made some educated guesses about which physical regimes could be associated with it. But is it a temporal process in any sense? Can we say, for example, that the universe underwent a transition to a geometric phase at some point in our past from an earlier non-geometric one? The issue if of course with respect to which temporal direction and variable we would make such statement. The question is very natural, since phase transitions provide prototypical examples of diachronic emergence, of novel behaviour appearing as a result of the actual temporal evolution of a system. Think for example of ice melting as we raise its temperature, by pumping heat into the system. 

Still, it should also be recalled that the typical way in which the phase structure of a physical system and its phase transitions are studied is in the context of equilibrium statistical mechanics and the renormalization group, and temporal evolution does not really play any role in them (by definition, there is no temporal evolution at equilibrium). The treatment of the same system out of equilibrium, within a formalism that allows to talk properly of evolution {\it across the phase transition}, is possible but more involved. To prove, using a realistic out-of-equilibrium description, that the phase transitions we normally describe with equilibrium statistical mechanics (thus as associated to different values of (coupling) constants), are in fact processes produced by the (quantum) dynamics of the system, is possible in many cases but in remains an open issue in general. As a result, one normally appeals to the fact that equilibrium statistical mechanics is  a very good approximation to the actual physical behaviour of dynamical systems whenever their evolution is not too fast, and relies on the possibility of external observers in the lab to adjust the conditions of the physical system and (slowly) tune the values of the parameters/couplings that characterize its description within equilibrium statistical mechanics. 

Three points need to be noted now, to properly appreciate the corresponding situation in quantum gravity. First, although the physics of phase transitions requires the microscopic theory to be properly understood, we are discussing an issue that could in principle we investigated at the continuum level, since we are dealing with the interpretation of continuum phases and their associated transitions. Second, a formal description of the phase diagram of quantum gravity models in a context akin to \lq equilibrium (quantum) statistical mechanics\rq as opposed to an \lq out-of-equilibrium\rq description is inevitable at the fundamental level, simply because at the fundamental level we do not have a preferred temporal direction that can be used to define temporal evolution\footnote{This absence makes of course challenging also to define the notion of \lq equilibrium\rq states in quantum gravity, lacking the usual definition and construction. However, this challenge can be met.}. Third, in general we cannot assume any hidden observer, external to the system (i.e. to the spacetime (region) under consideration), that could tune the parameters entering such formulation, thus \lq driving\rq the system to toward a phase transition; this is even more evident if we think of a geometrogenesis phase transition to be associated to the very early universe. On this basis, we need to conclude that geometrogenesis cannot be seen as an example of diachronic emergence but as another kind of synchronic one. Better still, since time simply disappears from the fundamental quantum gravity description of the world, geometrogenesis is another kind of {\it a-chronic emergence} as all the other kinds of emergence of time in quantum gravity that we have discussed above.

This is the conclusion that seems to be forced upon us at the fundamental level. But can geometrogenesis be given a temporal interpretation at least in an approximate, partial sense? We have discussed how this possibility can be realized in classical General Relativity and then in a quantum version of General Relativity. We have also seen how the same could be possible even if quantum gravity is based on more radically non-spatiotemporal entities, and gravitational physics in terms of continuum fields and spacetime is akin to its hydrodynamic description. 

Indeed, a viable strategy to give an approximate temporal description of geometrogenesis could be to extend to this context the same relational strategy for the definition of time that we have seen at play in the other contexts. First, we can certainly try to reformulate the effective continuum dynamics that is emergent from quantum gravity in a relational language, in terms of some internal \lq clock\rq variable (itself only an emergent quantity from some underlying subset of degrees of freedom of the theory). Using this, we can follow the evolution of the system toward the early universe. If at some point in this \lq backward evolution\rq we encounter either the phenomenological effects or the structural features that we can associated to a geometrogenesis phase transition, then we would be allowed to state that such phase transition has indeed taken place in the early universe, at the beginning of our cosmic history. This cosmological scenario is being investigated in detail, for example, in the context of group field theory condensate cosmology \cite{Oriti:2016qtz, Marchetti:2020umh, Marchetti:2020qsq}.

But can we then follow the same evolution {\it across} the transition? The answer depends on whether the relational reformulation of the quantum gravity dynamics remains valid across the transition. 

This consideration, and the following, applies also if the relational strategy is applied in the very initial formulation of the theory, recasting it in a \lq deparametrized\rq form before quantization. Such strategy, analogous to the one applied extensively in quantum General Relativity, has been investigated also in the (tensorial) group field theory approach. For example, one could think of relying on this deparametrized form for setting up a non-equilibrium description of the system, in terms of which tackling the issue of quantum gravity phase transitions as dynamical processes.

If the relational reformulation does remain valid across the transition, we have a geometrogenesis phase transition that, in a precise sense, can be given a temporal characterization, since it somehow spares time itself from disappearance\footnote{The situation would then be very close to the emergent universe scenario in string gas cosmology.}. In general, we should not expect this to be the case, though. We should rather expect that the very conditions that allow for a relational rewriting of the dynamics of the theory fail to be valid in such extreme conditions. The universe in its spatiotemporal description and time itself would dissolve at geometrogenesis\footnote{Also in the context of this discussion of quantum gravity phase transitions and geometrogenesis, we should recall and stress the remark we have made about the two possible perspectives on the new non-spatiotemporal entities of quantum gravity. If one deems them as mathematical tools only and maintains instead an ontology of continuum fields, the non-geometric and non-spatiotemporal phases would have no reason to be considered physical or philosophically interesting. The technical challenges for the quantum gravity approach (e.g. the need to study the continuum phase diagram of the theory, to identify the geometric phases, understand their physics and to characterize when one has a phase transition) would remain the same. But the conceptual ones would not. And the questions about whether the geometrogenesis is a physical, if not temporal, process would be meaningless, since the non-geometric phase would not be physical or \lq real\rq in any way.}. 
We would be left with the fundamental description of the same universe provided by the underlying non-spatiotemporal quantum gravity theory. 

As we wrote above for the hydrodynamic-like transition in quantum gravity, also in this case the temporal description of the process leading to the disappearance of time, that is, ceases to be valid exactly when such disappearance takes place. There is no diachronic emergence across the two descriptions. There can be, at best, a diachronic (i.e. in terms of temporal change) approach {\it toward the disappearance of time}, and a diachronic account {\it from the emergence of time onward}. Once more, this is where the analogy with ordinary atoms and fluids, that live {\it in spacetime}, fails and we are left in search of new intuitions.

\section{Concluding remarks}
The search for new intuitions about a timeless universe, one in which time is an emergent notion and its emergence has the complex, multifaceted nature that we have discussed, is an important part of the search for a new understanding of time in contemporary physics. It should be clear from our discussion that this search can only be successful if it is a joint effort of mathematicians, theoretical and experimental physicists, and philosophers, because the issues to be solved to achieve such understanding are conceptual as much as they are physical and mathematical. Any more solid grasp on the hidden richness of time, that we could only glimpse at this stage, will be an exciting reward for this collective effort. 


\bibliographystyle{unsrtnat}
\bibliography{TimeEmergenceBib}

\providecommand{\noopsort}[1]{}\providecommand{\singleletter}[1]{#1}%
\begin{thebibliography}{97}
\providecommand{\natexlab}[1]{#1}
\providecommand{\url}[1]{\texttt{#1}}
\expandafter\ifx\csname urlstyle\endcsname\relax
  \providecommand{\doi}[1]{doi: #1}\else
  \providecommand{\doi}{doi: \begingroup \urlstyle{rm}\Url}\fi

\bibitem[Oriti(2014{\natexlab{a}})]{Oriti:2013jga}
Daniele Oriti.
\newblock {Disappearance and emergence of space and time in quantum gravity}.
\newblock \emph{Stud. Hist. Phil. Sci. B}, 46:\penalty0 186--199,
  2014{\natexlab{a}}.
\newblock \doi{10.1016/j.shpsb.2013.10.006}.

\bibitem[Oriti(2020)]{Oriti:2018tym}
Daniele Oriti.
\newblock \emph{{The Bronstein hypercube of quantum gravity}}, pages 25--52.
\newblock 4 2020.
\newblock \doi{10.1017/9781108655705.003}.

\bibitem[Oriti(2018)]{Oriti:2018dsg}
Daniele Oriti.
\newblock {Levels of spacetime emergence in quantum gravity}.
\newblock 7 2018.

\bibitem[Oriti(2017)]{Oriti:2016acw}
Daniele Oriti.
\newblock {The universe as a quantum gravity condensate}.
\newblock \emph{Comptes Rendus Physique}, 18:\penalty0 235--245, 2017.
\newblock \doi{10.1016/j.crhy.2017.02.003}.

\bibitem[Rovelli(2021)]{Rovelli:2021elq}
Carlo Rovelli.
\newblock {The layers that build up the notion of time}.
\newblock 5 2021.

\bibitem[Frigg and Hartmann(2020)]{sep-models-science}
Roman Frigg and Stephan Hartmann.
\newblock {Models in Science}.
\newblock In Edward~N. Zalta, editor, \emph{The {Stanford} Encyclopedia of
  Philosophy}. Metaphysics Research Lab, Stanford University, {S}pring 2020
  edition, 2020.

\bibitem[Ladyman and Ross(2007)]{Ladyman2007-LADETM}
James Ladyman and Don Ross.
\newblock \emph{Every Thing Must Go: Metaphysics Naturalized}.
\newblock Oxford University Press, 2007.

\bibitem[Morganti and Tahko(2017)]{Morganti2017-TAHMNM}
Matteo Morganti and Tuomas~E. Tahko.
\newblock Moderately naturalistic metaphysics.
\newblock \emph{Synthese}, 194\penalty0 (7):\penalty0 2557--2580, 2017.
\newblock \doi{10.1007/s11229-016-1068-2}.

\bibitem[Butterfield and Isham(1999)]{Butterfield1999-BUTOTE}
Jeremy Butterfield and Chris Isham.
\newblock On the emergence of time in quantum gravity.
\newblock In Jeremy Butterfield, editor, \emph{The Arguments of Time}, pages
  111--168. Published for the British Academy by Oxford University Press, 1999.

\bibitem[Butterfield(2011{\natexlab{a}})]{Butterfield2011-BUTLID}
Jeremy Butterfield.
\newblock Less is different: Emergence and reduction reconciled.
\newblock \emph{Foundations of Physics}, 41\penalty0 (6):\penalty0 1065--1135,
  2011{\natexlab{a}}.
\newblock \doi{10.1007/s10701-010-9516-1}.

\bibitem[Butterfield(2011{\natexlab{b}})]{Butterfield2011-BUTERA}
Jeremy Butterfield.
\newblock Emergence, reduction and supervenience: A varied landscape.
\newblock \emph{Foundations of Physics}, 41\penalty0 (6):\penalty0 920--959,
  2011{\natexlab{b}}.
\newblock \doi{10.1007/s10701-011-9549-0}.

\bibitem[Bedau and Humphreys(2008)]{10.7551/mitpress/9780262026215.001.0001}
Mark~A. Bedau and Paul Humphreys.
\newblock \emph{{Emergence: Contemporary Readings in Philosophy and Science}}.
\newblock The MIT Press, 03 2008.
\newblock ISBN 9780262268011.
\newblock \doi{10.7551/mitpress/9780262026215.001.0001}.
\newblock URL \url{https://doi.org/10.7551/mitpress/9780262026215.001.0001}.

\bibitem[Butterfield and Bouatta(2011)]{pittphilsci8554}
Jeremy Butterfield and Nazim Bouatta.
\newblock Emergence and reduction combined in phase transitions, February 2011.
\newblock URL \url{http://philsci-archive.pitt.edu/8554/}.
\newblock Contribution to the Frontiers of Fundamental Physics (FFP 11)
  Conference Proceedings.

\bibitem[Batterman(2011)]{Batterman2011-BATESA}
Robert~W. Batterman.
\newblock Emergence, singularities, and symmetry breaking.
\newblock \emph{Foundations of Physics}, 41\penalty0 (6):\penalty0 1031--1050,
  2011.
\newblock \doi{10.1007/s10701-010-9493-4}.

\bibitem[Batterman(2004)]{Batterman2004-BATCPA}
Robert Batterman.
\newblock Critical phenomena and breaking drops: Infinite idealizations in
  physics.
\newblock \emph{Studies in History and Philosophy of Science Part B: Studies in
  History and Philosophy of Modern Physics}, 36\penalty0 (2):\penalty0
  225--244, 2004.
\newblock \doi{10.1016/j.shpsb.2004.05.004}.

\bibitem[Castellani(2000)]{Castellani2000-CASREA}
Elena Castellani.
\newblock Reductionism, emergence, and effective field theories.
\newblock \emph{Studies in History and Philosophy of Science Part B: Studies in
  History and Philosophy of Modern Physics}, 33\penalty0 (2):\penalty0
  251--267, 2000.
\newblock \doi{10.1016/s1355-2198(02)00003-5}.

\bibitem[Chalmers(2006)]{Chalmers2006-CHASAW}
David~J. Chalmers.
\newblock Strong and weak emergence.
\newblock In P.~Davies and P.~Clayton, editors, \emph{The Re-Emergence of
  Emergence: The Emergentist Hypothesis From Science to Religion}. Oxford
  University Press, 2006.

\bibitem[Rueger(2000)]{Rueger2000-RUEPED}
Alexander Rueger.
\newblock Physical emergence, diachronic and synchronic.
\newblock \emph{Synthese}, 124\penalty0 (3):\penalty0 297--322, 2000.
\newblock \doi{10.1023/A:1005249907425}.

\bibitem[Yates(2013)]{Yates2013-YATE}
David Yates.
\newblock Emergence.
\newblock In Hal Pashler, editor, \emph{Encyclopaedia of the Mind}, pages
  283--7. SAGE Reference, 2013.

\bibitem[Sartenaer(2018)]{Sartenaer2018-SARFE}
Olivier Sartenaer.
\newblock Flat emergence.
\newblock \emph{Pacific Philosophical Quarterly}, 99\penalty0 (S1):\penalty0
  225--250, 2018.
\newblock \doi{10.1111/papq.12233}.

\bibitem[Bishop and Ellis(2020)]{Bishop2020-BISCEO-3}
Robert~C. Bishop and George F.~R. Ellis.
\newblock Contextual emergence of physical properties.
\newblock \emph{Foundations of Physics}, 50\penalty0 (5):\penalty0 481--510,
  2020.
\newblock \doi{10.1007/s10701-020-00333-9}.

\bibitem[Huggett and Wuthrich(2021)]{Huggett2021-HUGOON-3}
Nick Huggett and Christian Wuthrich.
\newblock Out of nowhere: Introduction: The emergence of spacetime.
\newblock 2021.

\bibitem[Crowther(2020)]{Crowther2020-CROABS-4}
Karen Crowther.
\newblock As below, so before: \textquoteleft{}synchronic\textquoteright and
  \textquoteleft{}diachronic\textquoteright conceptions of spacetime emergence.
\newblock \emph{Synthese}, 198\penalty0 (8):\penalty0 7279--7307, 2020.
\newblock \doi{10.1007/s11229-019-02521-1}.

\bibitem[Baron(2019)]{Baron2019-BARTCC-15}
Sam Baron.
\newblock The curious case of spacetime emergence.
\newblock \emph{Philosophical Studies}, 177\penalty0 (8):\penalty0 2207--2226,
  2019.
\newblock \doi{10.1007/s11098-019-01306-z}.

\bibitem[Bihan and Linnemann(2019)]{LeBihan2019-LEBHWL}
Baptiste~Le Bihan and Niels~Siegbert Linnemann.
\newblock Have we lost spacetime on the way? narrowing the gap between general
  relativity and quantum gravity.
\newblock \emph{Studies in History and Philosophy of Science Part B: Studies in
  History and Philosophy of Modern Physics}, 65:\penalty0 112--121, 2019.
\newblock \doi{10.1016/j.shpsb.2018.10.010}.

\bibitem[Huggett and W\"uthrich(2018)]{Huggett2018-HUGTTE}
Nick Huggett and Christian W\"uthrich.
\newblock The (a)temporal emergence of spacetime.
\newblock \emph{Philosophy of Science}, 85\penalty0 (December):\penalty0
  1190--1203, 2018.
\newblock \doi{10.1086/699723}.

\bibitem[Bihan(2018)]{LeBihan2018-LEBSEI-2}
Baptiste~Le Bihan.
\newblock Space emergence in contemporary physics: Why we do not need
  fundamentality, layers of reality and emergence.
\newblock \emph{Disputatio}, 10\penalty0 (49):\penalty0 71--95, 2018.
\newblock \doi{10.2478/disp-2018-0004}.

\bibitem[Crowther(2014)]{Crowther2014-CROAOO}
Karen Crowther.
\newblock \emph{Appearing Out of Nowhere: The Emergence of Spacetime in Quantum
  Gravity}.
\newblock PhD thesis, University of Sydney, 2014.

\bibitem[Batterman(2006)]{Batterman2006-BATHVM}
Robert~W. Batterman.
\newblock Hydrodynamics versus molecular dynamics: Intertheory relations in
  condensed matter physics.
\newblock \emph{Philosophy of Science}, 73\penalty0 (5):\penalty0 888--904,
  2006.
\newblock \doi{10.1086/518747}.

\bibitem[Batterman(2013)]{Batterman2013-BATTTO-5}
Robert~W. Batterman.
\newblock The tyranny of scales.
\newblock In Robert~W. Batterman, editor, \emph{The Oxford handbook of
  philosophy of physics}, pages 255--286. Oxford University Press, 2013.

\bibitem[Pooley(2013)]{Pooley2013-POOSAR}
Oliver Pooley.
\newblock Substantivalist and relationalist approaches to spacetime.
\newblock In Robert Batterman, editor, \emph{The Oxford Handbook of Philosophy
  of Physics}. Oxford University Press, 2013.

\bibitem[Hoefer(1998)]{Hoefer1998-HOEAVR}
Carl Hoefer.
\newblock Absolute versus relational spacetime: For better or worse, the debate
  goes on.
\newblock \emph{British Journal for the Philosophy of Science}, 49\penalty0
  (3):\penalty0 451--467, 1998.
\newblock \doi{10.1093/bjps/49.3.451}.

\bibitem[Curiel(2014)]{Curiel2014-CUROTE-4}
Erik Curiel.
\newblock On the existence of spacetime structure.
\newblock \emph{British Journal for the Philosophy of Science}, page axw014,
  2014.
\newblock \doi{10.1093/bjps/axw014}.

\bibitem[Giulini(2007)]{Giulini:2006yg}
Domenico Giulini.
\newblock {Some remarks on the notions of general covariance and background
  independence}.
\newblock \emph{Lect. Notes Phys.}, 721:\penalty0 105--120, 2007.
\newblock \doi{10.1007/978-3-540-71117-9_6}.

\bibitem[Gaul and Rovelli(2000)]{Gaul:1999ys}
Marcus Gaul and Carlo Rovelli.
\newblock {Loop quantum gravity and the meaning of diffeomorphism invariance}.
\newblock \emph{Lect. Notes Phys.}, 541:\penalty0 277--324, 2000.

\bibitem[Rovelli(2011)]{Rovelli:2009ee}
Carlo Rovelli.
\newblock {'Forget time'}.
\newblock \emph{Found. Phys.}, 41:\penalty0 1475--1490, 2011.
\newblock \doi{10.1007/s10701-011-9561-4}.

\bibitem[Rovelli(1991)]{Rovelli:1990ph}
Carlo Rovelli.
\newblock {What Is Observable in Classical and Quantum Gravity?}
\newblock \emph{Class. Quant. Grav.}, 8:\penalty0 297--316, 1991.
\newblock \doi{10.1088/0264-9381/8/2/011}.

\bibitem[Marolf(1995)]{Marolf:1994nz}
Donald Marolf.
\newblock {Almost ideal clocks in quantum cosmology: A Brief derivation of
  time}.
\newblock \emph{Class. Quant. Grav.}, 12:\penalty0 2469--2486, 1995.
\newblock \doi{10.1088/0264-9381/12/10/007}.

\bibitem[Rovelli(2002)]{Rovelli:2001bz}
Carlo Rovelli.
\newblock {Partial observables}.
\newblock \emph{Phys. Rev. D}, 65:\penalty0 124013, 2002.
\newblock \doi{10.1103/PhysRevD.65.124013}.

\bibitem[Dittrich(2006)]{Dittrich:2005kc}
B.~Dittrich.
\newblock {Partial and complete observables for canonical general relativity}.
\newblock \emph{Class. Quant. Grav.}, 23:\penalty0 6155--6184, 2006.
\newblock \doi{10.1088/0264-9381/23/22/006}.

\bibitem[Husain and Pawlowski(2012)]{Husain:2011tk}
Viqar Husain and Tomasz Pawlowski.
\newblock {Time and a physical Hamiltonian for quantum gravity}.
\newblock \emph{Phys. Rev. Lett.}, 108:\penalty0 141301, 2012.
\newblock \doi{10.1103/PhysRevLett.108.141301}.

\bibitem[Thiemann(2006)]{Thiemann:2004wk}
Thomas Thiemann.
\newblock {Reduced phase space quantization and Dirac observables}.
\newblock \emph{Class. Quant. Grav.}, 23:\penalty0 1163--1180, 2006.
\newblock \doi{10.1088/0264-9381/23/4/006}.

\bibitem[Giesel and Thiemann(2015)]{Giesel:2012rb}
Kristina Giesel and Thomas Thiemann.
\newblock {Scalar Material Reference Systems and Loop Quantum Gravity}.
\newblock \emph{Class. Quant. Grav.}, 32:\penalty0 135015, 2015.
\newblock \doi{10.1088/0264-9381/32/13/135015}.

\bibitem[Giesel and Herzog(2018)]{Giesel:2017roz}
Kristina Giesel and Adrian Herzog.
\newblock {Gauge invariant canonical cosmological perturbation theory with
  geometrical clocks in extended phase-space \textemdash{} A review and
  applications}.
\newblock \emph{Int. J. Mod. Phys. D}, 27\penalty0 (08):\penalty0 1830005,
  2018.
\newblock \doi{10.1142/S0218271818300057}.

\bibitem[Kiefer(2013)]{Kiefer:2013jqa}
Claus Kiefer.
\newblock {Conceptual Problems in Quantum Gravity and Quantum Cosmology}.
\newblock \emph{ISRN Math. Phys.}, 2013:\penalty0 509316, 2013.
\newblock \doi{10.1155/2013/509316}.

\bibitem[Kiefer(2012)]{Kiefer:2012cdl}
Claus Kiefer.
\newblock {Quantum Gravity: Whence, Whither?}
\newblock In \emph{{Quantum Field Theory and Gravity: Conceptual and
  Mathematical Advances in the Search for a Unified Framework}}, pages 1--13,
  2012.
\newblock \doi{10.1007/978-3-0348-0043-3_1}.

\bibitem[Thiemann(2007)]{Thiemann:2007pyv}
Thomas Thiemann.
\newblock \emph{{Modern Canonical Quantum General Relativity}}.
\newblock Cambridge Monographs on Mathematical Physics. Cambridge University
  Press, 2007.
\newblock ISBN 978-0-511-75568-2, 978-0-521-84263-1.
\newblock \doi{10.1017/CBO9780511755682}.

\bibitem[Hoehn et~al.(2021)Hoehn, Smith, and Lock]{Hoehn:2019fsy}
Philipp~A. Hoehn, Alexander R.~H. Smith, and Maximilian P.~E. Lock.
\newblock {Trinity of relational quantum dynamics}.
\newblock \emph{Phys. Rev. D}, 104\penalty0 (6):\penalty0 066001, 2021.
\newblock \doi{10.1103/PhysRevD.104.066001}.

\bibitem[Castro-Ruiz et~al.(2020)Castro-Ruiz, Giacomini, Belenchia, and
  Brukner]{Castro-Ruiz:2019nnl}
Esteban Castro-Ruiz, Flaminia Giacomini, Alessio Belenchia, and Caslav Brukner.
\newblock {Quantum clocks and the temporal localisability of events in the
  presence of gravitating quantum systems}.
\newblock \emph{Nature Commun.}, 11\penalty0 (1):\penalty0 2672, 2020.
\newblock \doi{10.1038/s41467-020-16013-1}.

\bibitem[Giacomini(2021)]{Giacomini:2021gei}
Flaminia Giacomini.
\newblock {Spacetime Quantum Reference Frames and superpositions of proper
  times}.
\newblock \emph{Quantum}, 5:\penalty0 508, 2021.
\newblock \doi{10.22331/q-2021-07-22-508}.

\bibitem[Rovelli and Smolin(1995)]{Rovelli:1994ge}
Carlo Rovelli and Lee Smolin.
\newblock {Discreteness of area and volume in quantum gravity}.
\newblock \emph{Nucl. Phys. B}, 442:\penalty0 593--622, 1995.
\newblock \doi{10.1016/0550-3213(95)00150-Q}.
\newblock [Erratum: Nucl.Phys.B 456, 753--754 (1995)].

\bibitem[Isham(1993)]{Isham:1992ms}
C.~J. Isham.
\newblock {Canonical quantum gravity and the problem of time}.
\newblock \emph{NATO Sci. Ser. C}, 409:\penalty0 157--287, 1993.

\bibitem[Bojowald et~al.(2011)Bojowald, Hohn, and Tsobanjan]{Bojowald:2010qw}
Martin Bojowald, Philipp~A Hohn, and Artur Tsobanjan.
\newblock {Effective approach to the problem of time: general features and
  examples}.
\newblock \emph{Phys. Rev. D}, 83:\penalty0 125023, 2011.
\newblock \doi{10.1103/PhysRevD.83.125023}.

\bibitem[Marolf(2009)]{Marolf:2009wp}
Donald Marolf.
\newblock {Solving the Problem of Time in Mini-superspace: Measurement of Dirac
  Observables}.
\newblock \emph{Phys. Rev. D}, 79:\penalty0 084016, 2009.
\newblock \doi{10.1103/PhysRevD.79.084016}.

\bibitem[Alesci and Cianfrani(2014)]{Alesci:2014uha}
Emanuele Alesci and Francesco Cianfrani.
\newblock {Quantum Reduced Loop Gravity: Semiclassical limit}.
\newblock \emph{Phys. Rev. D}, 90\penalty0 (2):\penalty0 024006, 2014.
\newblock \doi{10.1103/PhysRevD.90.024006}.

\bibitem[Barrett et~al.(2011)Barrett, Dowdall, Fairbairn, Gomes, Hellmann, and
  Pereira]{Barrett:2010ex}
John~W. Barrett, R.~J. Dowdall, Winston~J. Fairbairn, Henrique Gomes, Frank
  Hellmann, and Roberto Pereira.
\newblock {Asymptotics of 4d spin foam models}.
\newblock \emph{Gen. Rel. Grav.}, 43:\penalty0 2421--2436, 2011.
\newblock \doi{10.1007/s10714-010-0983-7}.

\bibitem[Asante et~al.(2021)Asante, Dittrich, and
  Padua-Arguelles]{Asante:2021zzh}
Seth~K. Asante, Bianca Dittrich, and Jos\'e Padua-Arguelles.
\newblock {Effective spin foam models for Lorentzian quantum gravity}.
\newblock \emph{Class. Quant. Grav.}, 38\penalty0 (19):\penalty0 195002, 2021.
\newblock \doi{10.1088/1361-6382/ac1b44}.

\bibitem[Agullo and Singh(2017)]{Agullo:2016tjh}
Ivan Agullo and Parampreet Singh.
\newblock \emph{{Loop Quantum Cosmology}}, pages 183--240.
\newblock WSP, 2017.
\newblock \doi{10.1142/9789813220003_0007}.

\bibitem[Sahlmann(2010)]{Sahlmann:2010zf}
Hanno Sahlmann.
\newblock {Loop Quantum Gravity - A Short Review}.
\newblock In \emph{{Foundations of Space and Time: Reflections on Quantum
  Gravity}}, pages 185--210, 1 2010.

\bibitem[Perez(2013)]{Perez:2012wv}
Alejandro Perez.
\newblock {The Spin Foam Approach to Quantum Gravity}.
\newblock \emph{Living Rev. Rel.}, 16:\penalty0 3, 2013.
\newblock \doi{10.12942/lrr-2013-3}.

\bibitem[Oriti(2014{\natexlab{b}})]{Oriti:2014uga}
Daniele Oriti.
\newblock {Group Field Theory and Loop Quantum Gravity}.
\newblock 8 2014{\natexlab{b}}.

\bibitem[Loll(2020)]{Loll:2019rdj}
R.~Loll.
\newblock {Quantum Gravity from Causal Dynamical Triangulations: A Review}.
\newblock \emph{Class. Quant. Grav.}, 37\penalty0 (1):\penalty0 013002, 2020.
\newblock \doi{10.1088/1361-6382/ab57c7}.

\bibitem[Gurau and Ryan(2012)]{Gurau:2012hl}
Razvan Gurau and James~P Ryan.
\newblock {Colored Tensor Models - a Review}.
\newblock \emph{SIGMA}, 8:\penalty0 020, 2012.

\bibitem[Oriti(2011)]{Oriti:2011jm}
Daniele Oriti.
\newblock {The microscopic dynamics of quantum space as a group field theory}.
\newblock In \emph{{Foundations of Space and Time: Reflections on Quantum
  Gravity}}, pages 257--320, 10 2011.

\bibitem[Krajewski(2011)]{Krajewski:2012aw}
Thomas Krajewski.
\newblock {Group field theories}.
\newblock \emph{PoS}, QGQGS2011:\penalty0 005, 2011.
\newblock \doi{10.22323/1.140.0005}.

\bibitem[Surya(2019)]{Surya:2019ndm}
Sumati Surya.
\newblock {The causal set approach to quantum gravity}.
\newblock \emph{Living Rev. Rel.}, 22\penalty0 (1):\penalty0 5, 2019.
\newblock \doi{10.1007/s41114-019-0023-1}.

\bibitem[Blau and Theisen(2009)]{Blau:1900zza}
Matthias Blau and Stefan Theisen.
\newblock {String theory as a theory of quantum gravity: A status report}.
\newblock \emph{Gen. Rel. Grav.}, 41:\penalty0 743--755, 2009.
\newblock \doi{10.1007/s10714-008-0752-z}.

\bibitem[Aoki et~al.(1998)Aoki, Iso, Kawai, Kitazawa, and Tada]{Aoki:1998vn}
Hajime Aoki, Satoshi Iso, Hikaru Kawai, Yoshihisa Kitazawa, and Tsukasa Tada.
\newblock {Space-time structures from IIB matrix model}.
\newblock \emph{Prog. Theor. Phys.}, 99:\penalty0 713--746, 1998.
\newblock \doi{10.1143/PTP.99.713}.

\bibitem[Bahr and Steinhaus(2017)]{Bahr:2017klw}
Benjamin Bahr and Sebastian Steinhaus.
\newblock {Hypercuboidal renormalization in spin foam quantum gravity}.
\newblock \emph{Phys. Rev.}, D95\penalty0 (12):\penalty0 126006, 2017.
\newblock \doi{10.1103/PhysRevD.95.126006}.

\bibitem[Dittrich(2017)]{Dittrich:2014ala}
Bianca Dittrich.
\newblock {The continuum limit of loop quantum gravity - a framework for
  solving the theory}.
\newblock In Abhay Ashtekar and Jorge Pullin, editors, \emph{Loop Quantum
  Gravity: The First 30 Years}, pages 153--179. 2017.
\newblock \doi{10.1142/9789813220003_0006}.

\bibitem[Eichhorn et~al.(2019)Eichhorn, Koslowski, and
  Pereira]{Eichhorn:2018phj}
Astrid Eichhorn, Tim Koslowski, and Antonio~D. Pereira.
\newblock {Status of background-independent coarse-graining in tensor models
  for quantum gravity}.
\newblock \emph{Universe}, 5\penalty0 (2):\penalty0 53, 2019.
\newblock \doi{10.3390/universe5020053}.

\bibitem[Carrozza(2016)]{Carrozza:2016vsq}
Sylvain Carrozza.
\newblock {Flowing in Group Field Theory Space: a Review}.
\newblock \emph{SIGMA}, 12:\penalty0 070, 2016.
\newblock \doi{10.3842/SIGMA.2016.070}.

\bibitem[Finocchiaro and Oriti(2021)]{Finocchiaro:2020fhl}
Marco Finocchiaro and Daniele Oriti.
\newblock {Renormalization of group field theories for quantum gravity: new
  scaling results and some suggestions}.
\newblock \emph{Front. in Phys.}, 8:\penalty0 649, 2021.
\newblock \doi{10.3389/fphy.2020.552354}.

\bibitem[Pithis and Sakellariadou(2019)]{Pithis:2019tvp}
A.G.A. Pithis and M.~Sakellariadou.
\newblock {Group field theory condensate cosmology: An appetizer}.
\newblock \emph{Universe}, 5\penalty0 (6):\penalty0 147, 2019.
\newblock \doi{10.3390/universe5060147}.

\bibitem[Gielen and Sindoni(2016)]{Gielen:2016dss}
Steffen Gielen and Lorenzo Sindoni.
\newblock {Quantum Cosmology from Group Field Theory Condensates: a Review}.
\newblock \emph{SIGMA}, 12:\penalty0 082, 2016.
\newblock \doi{10.3842/SIGMA.2016.082}.

\bibitem[Oriti et~al.(2016)Oriti, Sindoni, and Wilson-Ewing]{Oriti:2016qtz}
Daniele Oriti, Lorenzo Sindoni, and Edward Wilson-Ewing.
\newblock {Emergent Friedmann dynamics with a quantum bounce from quantum
  gravity condensates}.
\newblock \emph{Class. Quant. Grav.}, 33\penalty0 (22):\penalty0 224001, 2016.
\newblock \doi{10.1088/0264-9381/33/22/224001}.

\bibitem[Marchetti and Oriti(2021{\natexlab{a}})]{Marchetti:2020umh}
Luca Marchetti and Daniele Oriti.
\newblock {Effective relational cosmological dynamics from Quantum Gravity}.
\newblock \emph{JHEP}, 05:\penalty0 025, 2021{\natexlab{a}}.
\newblock \doi{10.1007/JHEP05(2021)025}.

\bibitem[Oriti and Pang(2021)]{Oriti:2021rvm}
Daniele Oriti and Xiankai Pang.
\newblock {Phantom-like dark energy from quantum gravity}.
\newblock 5 2021.

\bibitem[Wilson-Ewing(2019)]{Wilson-Ewing:2018mrp}
Edward Wilson-Ewing.
\newblock {A relational Hamiltonian for group field theory}.
\newblock \emph{Phys. Rev.}, D99\penalty0 (8):\penalty0 086017, 2019.
\newblock \doi{10.1103/PhysRevD.99.086017}.

\bibitem[Gielen and Polaczek(2020)]{Gielen:2019kae}
Steffen Gielen and Axel Polaczek.
\newblock {Generalised effective cosmology from group field theory}.
\newblock \emph{Class. Quant. Grav.}, 37\penalty0 (16):\penalty0 165004, 2020.
\newblock \doi{10.1088/1361-6382/ab8f67}.

\bibitem[Gielen(2021)]{Gielen:2021dlk}
Steffen Gielen.
\newblock {Frozen formalism and canonical quantization in (group) field
  theory}.
\newblock 5 2021.

\bibitem[Delcamp and Dittrich(2017)]{Delcamp:2016dqo}
Clement Delcamp and Bianca Dittrich.
\newblock {Towards a phase diagram for spin foams}.
\newblock \emph{Class. Quant. Grav.}, 34\penalty0 (22):\penalty0 225006, 2017.
\newblock \doi{10.1088/1361-6382/aa8f24}.

\bibitem[Kegeles et~al.(2018)Kegeles, Oriti, and Tomlin]{Kegeles:2017ems}
Alexander Kegeles, Daniele Oriti, and Casey Tomlin.
\newblock {Inequivalent coherent state representations in group field theory}.
\newblock \emph{Class. Quant. Grav.}, 35\penalty0 (12):\penalty0 125011, 2018.
\newblock \doi{10.1088/1361-6382/aac39f}.

\bibitem[Bahr et~al.(2021)Bahr, Dittrich, and Geiller]{Bahr:2015bra}
Benjamin Bahr, Bianca Dittrich, and Marc Geiller.
\newblock {A new realization of quantum geometry}.
\newblock \emph{Class. Quant. Grav.}, 38\penalty0 (14):\penalty0 145021, 2021.
\newblock \doi{10.1088/1361-6382/abfed1}.

\bibitem[Koslowski and Sahlmann(2012)]{Koslowski:2011vn}
Tim Koslowski and Hanno Sahlmann.
\newblock {Loop quantum gravity vacuum with nondegenerate geometry}.
\newblock \emph{SIGMA}, 8:\penalty0 026, 2012.
\newblock \doi{10.3842/SIGMA.2012.026}.

\bibitem[Surya(2012)]{Surya:2011du}
Sumati Surya.
\newblock {Evidence for a Phase Transition in 2D Causal Set Quantum Gravity}.
\newblock \emph{Class. Quant. Grav.}, 29:\penalty0 132001, 2012.
\newblock \doi{10.1088/0264-9381/29/13/132001}.

\bibitem[Pithis and Th\"urigen(2020)]{Pithis:2020kio}
Andreas G.~A. Pithis and Johannes Th\"urigen.
\newblock {Phase transitions in TGFT: functional renormalization group in the
  cyclic-melonic potential approximation and equivalence to O$(N)$ models}.
\newblock \emph{JHEP}, 12:\penalty0 159, 2020.
\newblock \doi{10.1007/JHEP12(2020)159}.

\bibitem[Ambj\o{}rn et~al.(2017)Ambj\o{}rn, Gizbert-Studnicki, G\"orlich,
  Jurkiewicz, Klitgaard, and Loll]{Ambjorn:2016mnn}
J.~Ambj\o{}rn, J.~Gizbert-Studnicki, A.~G\"orlich, J.~Jurkiewicz, N.~Klitgaard,
  and R.~Loll.
\newblock {Characteristics of the new phase in CDT}.
\newblock \emph{Eur. Phys. J. C}, 77\penalty0 (3):\penalty0 152, 2017.
\newblock \doi{10.1140/epjc/s10052-017-4710-3}.

\bibitem[Konopka et~al.(2006)Konopka, Markopoulou, and Smolin]{Konopka:2006hu}
Tomasz Konopka, Fotini Markopoulou, and Lee Smolin.
\newblock {Quantum Graphity}.
\newblock 11 2006.

\bibitem[Oriti(2007)]{Oriti:2007qd}
Daniele Oriti.
\newblock {Group field theory as the microscopic description of the quantum
  spacetime fluid: A New perspective on the continuum in quantum gravity}.
\newblock \emph{PoS}, QG-PH:\penalty0 030, 2007.

\bibitem[Mandrysz and Mielczarek(2019)]{Mandrysz:2018sle}
Micha~L. Mandrysz and Jakub Mielczarek.
\newblock {Ultralocal nature of geometrogenesis}.
\newblock \emph{Class. Quant. Grav.}, 36\penalty0 (1):\penalty0 015004, 2019.
\newblock \doi{10.1088/1361-6382/aaef71}.

\bibitem[Smolin(2003)]{Smolin:2003qf}
L.~Smolin.
\newblock {The self-organization of space and time}.
\newblock \emph{Phil. Trans. Roy. Soc. Lond. A}, 361:\penalty0 1081--1088,
  2003.
\newblock \doi{10.1098/rsta.2003.1185}.

\bibitem[Percacci(1991)]{Percacci:1990wy}
R.~Percacci.
\newblock {The Higgs phenomenon in quantum gravity}.
\newblock \emph{Nucl. Phys. B}, 353:\penalty0 271--290, 1991.
\newblock \doi{10.1016/0550-3213(91)90510-5}.

\bibitem[Brandenberger(2015)]{Brandenberger:2015kga}
Robert~H. Brandenberger.
\newblock {String Gas Cosmology after Planck}.
\newblock \emph{Class. Quant. Grav.}, 32\penalty0 (23):\penalty0 234002, 2015.
\newblock \doi{10.1088/0264-9381/32/23/234002}.

\bibitem[Magueijo et~al.(2007)Magueijo, Smolin, and Contaldi]{Magueijo:2006fu}
Joao Magueijo, Lee Smolin, and Carlo~R. Contaldi.
\newblock {Holography and the scale-invariance of density fluctuations}.
\newblock \emph{Class. Quant. Grav.}, 24:\penalty0 3691--3700, 2007.
\newblock \doi{10.1088/0264-9381/24/14/009}.

\bibitem[Mielczarek(2017)]{Mielczarek:2014aka}
Jakub Mielczarek.
\newblock {Big Bang as a critical point}.
\newblock \emph{Adv. High Energy Phys.}, 2017:\penalty0 4015145, 2017.
\newblock \doi{10.1155/2017/4015145}.

\bibitem[Marchetti and Oriti(2021{\natexlab{b}})]{Marchetti:2020qsq}
Luca Marchetti and Daniele Oriti.
\newblock {Quantum fluctuations in the effective relational GFT cosmology}.
\newblock \emph{Front. Astron. Space Sci.}, 8:\penalty0 683649,
  2021{\natexlab{b}}.
\newblock \doi{10.3389/fspas.2021.683649}.

\end{thebibliography}

\end{document}